\documentstyle[eqsecnum,aps,epsf,prl]{revtex}

\begin{document}
\title{Strong and electromagnetic interaction views on deuteron structure.}
\author{Egle Tomasi-Gustafsson}
\address{ DAPNIA/SPhN, CEA/Saclay, 91191 Gif-sur-Yvette Cedex, 
France}
\author{Michail P. Rekalo \footnote{ Permanent address:
\it National Science Center KFTI, 310108 Kharkov, Ukraine}
}
\address{Middle East Technical University, 
Physics Department, Ankara 06531, Turkey}
\maketitle
\date{\today}
\def\gms{$G_{Ms}$}
\def\gmp{$G_{Mp}$}
\def\gmn{$G_{Mn}$}
\def\ges{$G_{Es}$}
\def\gep{$G_{Ep}$}
\def\gen{$G_{En}$}
\begin{abstract}
The purpose of this contribution is to give an outlook of recent results 
connected with deuteron physics, with electromagnetic and strong interacting 
probes at intermediate energy. Special attention will be devoted to 
polarization 
observables.
\end{abstract}
\section{Introduction}

The deuteron (bound state of proton and neutron) with spin ${\cal J}$ and 
parity $P$: ${\cal J}^P=1^+$, and with isospin $I$=0, has been investigated, 
theoretically and experimentally, since many decades. Few selected problems, 
which are very actual, will be discussed here according the 
following plan: 
\begin{itemize}
\item  The knowledge of the deuteron structure, itself, at the largest possible 
internal momenta (or at the shortest distances between 
the nucleons). The question which is 
addressed is the definition of the kinematical region where a description 
based on nucleons and mesons (basically impulse approximation (IA) with 
corrections due to meson exchange currents, isobars, relativistic effects..) 
has to be 
discarded in favor of a 'high energy view' in terms of quarks and gluons. Here, 
in particular, pQCD gives predictions concerning the asymptotic behavior of the 
deuteron form factors and assumes hadron helicity conservation for the 
amplitudes.

The transition region between these two regimes should be the privileged domain 
of 
intermediate energy machines. High intensity is required by exclusive 
measurements and/or studies with secondary beams (polarization etc..).

\item 
The deuteron as a probe to investigate the nucleon or 
the heavy nuclei structure.

\begin{itemize}

\item it can be used to study the properties of the neutron, which is not 
available as a target;

\item it is an isoscalar probe, which can be very selective in exciting 
specific states in nucleons and nuclei.
 
\end{itemize}
\end{itemize}
In order to illustrate the 
different points listed above, we will review some of the last data obtained at 
Jefferson Laboratory and 
Saturne. We will show few results from elastic electron-deuteron scattering
and proton-deuteron elastic and inelastic scattering. Particular attention  
will 
be devoted to polarization observables.

\section{The deuteron electromagnetic form factors}

\subsection{Recent determination of the elastic deuteron electromagnetic form 
factors}

The measurement of the differential cross section of elastic $ed-$scattering, 
for a fixed value of $Q^2$, 
at 
different scattering angles, allows to determine the structure functions 
$A(Q^2)$ 
and $B(Q^2)$:
$$\displaystyle\frac{d\sigma}{d\Omega}=
\left (\displaystyle\frac{d\sigma}{d\Omega}\right)_0\cdot
{\cal S},~~{\cal S}= A(Q^2)+B(Q^2) ~\tan^2(\theta_e/2)
$$
with
$$
\left (\displaystyle\frac{d\sigma}{d\Omega}\right)_0=
\frac{\alpha^2~cos^2(\theta_e/2)E'}{4E^3sin^4(\theta_e/2)},
$$
where $E$ ($E'$)  is the electron beam (the scattered electron) energy and 
$\theta_e$ the electron scattering angle in the Laboratory system. 
The structure functions $A$ and $B$ can be expressed in terms of the three form 
factors, $G_c$ 
(electric), $G_m$ (magnetic) and $G_q$ (quadrupole) as:
$$A(Q^2) =G_c^2(Q^2)+ \frac{8}{9} \tau^2 G_Q^2(Q^2)+\frac{2}{3} \tau 
G_m^2(Q^2),~~B(Q^2) = 
\frac{4}{3} (1+\tau) \tau G_m^2(Q^2),~~\tau=\frac{Q^2}{4M^2},
$$
where M is the deuteron mass. In case of unpolarized beam and target 
the outgoing deuteron is tensorially polarized and the components of the tensor 
polarization give useful combinations of form factors. In particular $t_{20}$ 
allows, together with $A(Q^2)$ and $B(Q^2)$, the 
determination of the three form factors: 
$$
t_{20}=-\displaystyle\frac{1}{\sqrt{2}{\cal S}}\left [
\displaystyle\frac{8}{3}\tau G_c G_q + \displaystyle\frac{8}{9} \tau^2 
G_Q^2+\frac{1}{3} \tau
\left (1+2(1+\tau)\tan^2 (\theta_e/2)\right )\right ]G_m^2,
$$
At the Jefferson Laboratory (JLab), the elastic $ed-$ cross section has been 
recently precisely measured up to large momentum transfer $Q^2\simeq 6$ 
(GeV/c)$^{-2}$, 
\cite{Al99,Ab99} and $t_{20}$ has been measured for  
a momentum 
transfer up to $Q^2=1.9$ (GeV/c)$^{-2}$  \cite{Ab00}. 

According to \cite{Al99}, the cross sections seems to scale as $(Q^2)^{-10}$, as 
previously pointed out \cite{Ar75}, and predicted by pQCD.
However, from the $t_{20}$ data, it clearly appears that the pQCD 
limit is not 
yet reached, and that the data follow the trend suggested by IA. On the other 
hand, it is not possible, from these data, to constrain definitely different 
models or corrections. For a detailed comparison with theory, see, 
for example, \cite{Al99,Ab00}.

The question is then, how to proceed further.
On the experimental point of view, 
in the best presently achievable conditions of luminosity, about one month 
beam time was needed 
for the  the  polarization  measurement  and 
one week for the cross section at the largest values of $Q^2$. It 
seems hard to foresee more favorable experimental conditions, with the present 
technology related to electron machines.
One possibility is to investigate the inelastic deuteron form factors through 
the reaction $ed\to ed\pi^0$, where for the same momentum transfer for the 
electron, one acceeds to shortest distances in the deuteron, compared to 
elastic 
scattering \cite{Re00}.

In next section we will focus on a conceptual problem,  which is the limit of 
the validity of the assumption of the one 
photon exchange mechanism in electron-hadron scattering at large momentum 
transfer.  A 
related problem, that we will not discuss here, is that  
radiative corrections for polarization effects at large momentum transfer are 
not known.

\subsection{Beyond the one-photon approximation}

The formulas given above are valid if the momentum is transferred from the 
incident electron to the target by a virtual photon, with the underlying 
assumption that the possible two-photon contribution is small. 
The relative contribution of two photon exchange, from simple counting in 
$\alpha$, would be of the order of the fine 
structure constant, $\alpha=\displaystyle\frac{e^2}{4\pi}\simeq \frac{1}{137}$. 
However, more than 25 years ago it was observed that 
the relative role of two-photon exchange can increase significantly in the 
region 
of high momentum transfer \cite{Gu73,Fr73,Bo73,Le73}. If the transferred 
momentum is equally shared between two virtual photons, due to the steep 
decrease of the deuteron form factors, the simple rule of $\alpha$-counting for 
the estimation of the relative 
role of two-photon contribution to the amplitude of elastic $ed-$scattering  
does not hold anymore. This effect would  manifest already at  
momentum transfer of the order of 1 GeV$^2$, in particular  in 
the region of  diffractive minima.
 
In Ref. \cite{Gu73} the two-photon amplitude is purely imaginary, at least at 
small 
scattering angles, so it cannot interfere with the one-photon exchange 
amplitude 
in the differential cross section for unpolarized particles scattering.
However, in this case, the polarization observables in elastic 
$ed-$scattering have to be large, in particular the T-odd polarization 
observables. But the predicted increasing of the 
two-photon mechanism is so large that it may be observed even in the 
differential 
cross section of elastic $ed-$scattering, at relatively large momentum transfer 
square, $Q=8-12$ fm$^{-1}$. An evaluation of this contribution from the existing
experimental data has been done in \cite{Re99}.

The crossing symmetry 
can provide a relation between the matrix elements ${\cal M}$ of the  
the elastic $e^-+h\rightarrow e^-+h$ 
scattering  and the $e^+e^-$-annihilation: 
$e^++e^-\rightarrow  \overline{h}+h$, in one-photon approximation.
\begin{equation}
\overline{|{\cal M}(eh\rightarrow eh)|^2}=f(s,t)=\overline{|{\cal 
M}(e^+e^-\rightarrow \overline h h)|^2}.
\end{equation}
The line over ${\cal M}$ denotes the sum over the polarizations of all 
particles (in initial and final states). The Mandelstam variable $s$ is the 
total 
energy square  and $t$ is  the momentum transfer square. They delimit different 
kinematical 
regions 
for the annihilation  and the 
scattering channel.

The presence of a single virtual photon in the reaction 
$e^++e^-\rightarrow\gamma^*\rightarrow\overline{h}+{h}$ constrains the
total angular momentum ${\cal J}$ and the $P$-parity 
for the $\overline{h}{h}-$system  to take only one 
possible value, ${\cal J}^P=1^-$, the quantum number of the photon.  
In 
the framework of the one-photon approximation, in the general case,
 $\overline{|{\cal M}(e^+e^- \rightarrow h\overline{h})|^2}$ 
can be written (in CMS)  as:

\begin{equation}
\overline{|{\cal M}(e^+e^\rightarrow h\overline{h})|^2}
=a(t)+b(t)\cos^2\tilde\theta,~\mbox{and~}
\cos^2\tilde\theta=1+\frac{\cot^2\frac{\theta_e}{2}}{1+\tau},
\label{eq:cse}
\end{equation}
where $a(t)$ and $b(t)$ are definite quadratic combinations of the 
electromagnetic form factors for the hadron $h$ and $\tilde\theta$ is the angle 
of the detected hadron. 

In case of the presence of $2\gamma$ in the intermediate state, in the 
annihilation channel,  any value of 
the total angular momentum and space parity is allowed, because 
the relative 3-momentum for 
the $2\gamma$-state is nonzero, contrary to the case  of the one-photon 
mechanism.  The $\overline{h}h$-system, produced 
through $1\gamma$- and $2\gamma$-exchanges has different values of C-parity, 
because $C(\gamma)=-1$ and $C(2\gamma)=+1$. Therefore the 
interference of  one- and two-photon contribution must be an {\bf odd} function 
of $\cos\tilde\theta$: 
$\overline{{\cal R}e{\cal M}_1{\cal 
M}_2^*}=\cos\tilde\theta(a_0+a_2\cos^2\tilde\theta +...)$.  

The SLAC results \cite{Ar75} where obtained at a fixed electron scattering 
angle.
The odd contribution can then be estimated from the cross sections at two 
angles, for the same $Q$, one value being given from a fit of the data of ref. 
\cite{Ar75} and the other by 
the recent data  \cite{Al99,Ab99}. We can then calculate separately the linear 
contribution 
or the cubic contribution in $\cos\tilde\theta$. It is expected that deviation 
from the linear 
$\cot^2\frac{\theta_e}{2}$ formula 
would not appear for $Q \le$ 5 fm$^{-1}$.
The resulting ratios C/A and D/A  are reported in Fig. 1 and 2  as  functions 
of $Q$, as open circles \cite{Al99} and open squares \cite{Ab99}. The data 
nicely agree: at large momentum transfer  these ratios deviate from 
zero and show a dependence on the transferred momentum, 
which could result from two photon exchange. However the points had to be 
rescaled in order to have zero deviation at low $Q$ 
(corresponding solid symbols) due to  systematic errors in the 
measurement of the cross section between the two sets of data \cite{Al99} and 
\cite{Ab99}.

While this can  not be considered as a definite evidence for the presence of 
$2\gamma$-exchange in $ed-$elastic scattering, it is  the first attempt 
to obtain 
a quantitative upper limit of a possible $2\gamma$-contribution, using a 
parameterization of the $1\gamma\bigotimes 2\gamma$-interference and the 
existing experimental data.

The $2\gamma$-exchange in elastic hadron scattering can be experimentally 
searched
in different ways: - through the comparison of the cross section for scattering 
of unpolarized electrons and 
positrons (by protons or deuterons) in the same kinematical conditions, - 
looking to the deviation from a straight line on the Rosenbluth plot or  
measuring specific properties of polarization phenomena: as nonzero  $T-odd$ 
polarization observables, and violation of definite relations between T-even 
polarization observables
and the SF $B(Q^2)$.
The measurement of cross section and polarization observables bring 
complementary 
and independent pieces of information, as they test the real and imaginary part 
of the $2\gamma$ contribution.

\section{The deuteron structure from $dp$ break-up and backward elastic 
scattering}

We showed that the measurement of cross section and analyzing powers in 
$ed-$elastic scattering  allows to determine all the three form factors (in 
one-photon approximation). These form factors, in IA, are integrals of the wave 
function over the radial coordinate. On the other hand, the reactions $ \vec 
d+p\to \vec p+d$ (backward elastic scattering) and $\vec d+p\to p+X$ (deuteron 
break up) are directly 
related to the deuteron wave functions. In the IA, the tensor analyzing power 
$T_{20}$ and the polarization transfer $\kappa_0$ can be written as:

$$T_{20}=\displaystyle\frac{1}{\sqrt{2}}\displaystyle\frac{\sqrt{8}uw-w^2}{u^2+
w^2},$$
$$\kappa_0=\displaystyle\frac{u^2-w^2-{1}{2}uw}{u^2+w^2},$$
from which one obtains a quadratic relation between $T_{20}$ and $\kappa_0$ :
$$ \left (T_{20}+\displaystyle\frac{1}{2\sqrt{2}}\right 
)^2+\kappa_0^2=\displaystyle\frac{9}{8}$$
Measurements have been performed at Saturne and Dubna \cite{cfp,dubna}. The 
correlation 
between the two polarization observables is shown in Fig. 3.
The full line is the IA prediction. The two sets of points correspond to 
backward 
elastic scattering (open circles) and deuteron break up (solid circles) and 
they 
show a very similar behavior. The deviation from IA,
which gets larger at larger momenta has been interpreted in different models 
(see \cite{Kuehn} and refs. herein). At this moment it seems fair to conclude 
that unambiguous 
signatures of quarks have not yet been found. 

A promising way of looking to deuteron at very short distances seems to be the 
measurement of the tensor analyzing power of pions emitted at $0^0$, in the 
reaction $\vec d +p\to \pi+X$, in the 
cumulative region \cite{Li00}. 

\section{ The deuteron as an isoscalar probe}

\subsection{The neutron electromagnetic form factors}

Having high precision data on the differential cross section for $ed-$ elastic 
scattering, and assuming a 
reliable model for their description,  one can extract, in principle, the 
dependence of the electric neutron form factor \gen\  on the momentum transfer 
$Q^2$. Such a procedure  has been carried out in ref. \cite{Pl90}, up to 
$Q^2$=0.7 (GeV/c)$^2$. It can be extended at higher $Q^2$ using the  elastic 
$ed$-scattering data mentioned above and recent data on the proton electric 
form factor 
\cite{Jo00}. These data have been obtain by the recoil proton polarization 
measurement in $\vec e+p\to e+\vec p$, following an idea suggested more than 30 
years ago \cite{Re68} and extend up to $Q^2$=3.5 
(GeV/c)$^2$.
The large sensitivity to 
the nucleon form factors of the models which 
describe the light nuclei structure, particularly the deuteron, was already 
carefully studied in \cite{Ar80}, and 
it was pointed out that the disagreement between the relativistic impulse 
approximation and the data  could be 
significantly reduced if \gen\  were different from zero. 

In the non relativistic IA, the deuteron form factors
depend only on the deuteron wave function and on nucleon form factors:
\begin{equation}
G_c=G_{Es}C_E,~~G_q=G_{Es}C_Q,~~G_m=\displaystyle\frac{M_d}{M_p}\left 
(G_{Ms}C_S+\displaystyle\frac{1}{2}G_{Es}C_L\right ),
\end{equation}
where $M_p$ is the proton mass, \ges=\gep+\gen\  and \gms=\gmp+\gmn\  are the 
charge and 
magnetic isoscalar nucleon form factors, respectively. The terms 
$C_E$, $C_Q$, 
$C_S$, and $C_L$ describe the deuteron structure and can be calculated from the 
deuteron $S$ and $D$ 
wave functions, $u(r)$ and $w(r)$ \cite{Ja56} :
$$C_E={\int }_0^{\infty}dr~j_0\left( 
\frac{Qr}2\right) \left[ u^2\left( r\right) +w^2( r)
\right], $$
$$C_Q=\frac{3}{\sqrt{2}\eta}{\int }_0^{\infty}dr~j_2\left( 
\frac{Qr}2\right) \left[ u( r) -\frac{w( r)}{\sqrt{8}}\right] w(r),  $$
\begin{equation}
C_S={\int }_0^{\infty}dr \left[ u^2( r) -\frac{1}2w^2( r)
\right ]j_0\left( \frac{Qr}{2}\right) + 
\frac{1}{2}\left [\sqrt{2}u( r)w(r)+w^2( r)\right ] j_2\left( 
\frac{Qr}{2} \right ),
\end{equation}
$$C_L=\frac{3}{2} \int_0^{\infty}dr~w^2( r)
\left [ j_0 \left ( \frac{Qr}{2} \right )+ j_2 \left ( \frac{Qr}{2}\right ) 
\right ],
$$
where 
$j_0(x)$ and $ j_2( x)$
are the spherical Bessel functions.
The normalization condition is $
{\int }_0^{\infty}dr~\left[ u^2( r)+w^2( r)\right ]=1.$

With the help of expressions (3) and (4), the formula for $A(Q)^2$,  
can be inverted into a quadratic equation for \ges. Then \ges\  is calculated 
using the experimental values for $A(Q)^2$, assuming, for the magnetic nucleon 
form factors $G_{Mp}$ and $G_{Mn}$ 
the usual dipole dependence, which is in agreement with the existing data at a 
3\%  level, up to
$Q^2\simeq$ 10 (GeV/c)$^2$.

In Fig. 4 we illustrate the behavior of the different nucleon 
electric form factors: \ges,\  \gep\  and \gen. The nucleon isoscalar 
electric form factor, 
derived from different sets of deuteron data, decreases when $Q^2$ increases. 
The solid line represents the 
Gari-Kr\"umpelmann parametrization \cite{G-K} for \ges . The dipole behavior, 
which is 
generally 
assumed for the proton electric form factor is shown as a dotted line. 
The new \gep\  data, which decrease faster than the dipole function, are also 
well 
reproduced by the Gari-Kr\"umpelmann parametrization (thick dashed line). 

The  electric neutron form factor can be calculated 
from the isoscalar nucleon form factor, taking for \gep\  a dipole behavior 
(solid stars) or a fit based on the new data (open stars). The last option 
leads 
to values for 
\gen\  which are in  
very good agreement with the parametrization 
\cite{G-K}. These results shows that the neutron form factor
is not going to vanish identically at large momentum transfer, but becomes more 
sizeable 
than predicted by other parametrizations, often used in the calculations 
\cite{Pl90,Galster} (thin dashed line).
Starting from $Q^2\simeq 2$ (GeV/c)$^2$  the form factor \gen\  becomes even 
larger than \gep . Let us mention that a recent 'direct' measurement 
\cite{Ro99} 
at 
$Q^2=0.67$ (GeV/c)$^2$ finds \gen =$0.052\pm 0.011\pm 0.005$  in agreement with 
the present values.

Let us mention that the $\gamma^*\pi^{\pm}\rho^{\mp}$-contribution, which is a 
good approximation for the isoscalar 
transition $\gamma^*\rightarrow \pi^+\pi^-\pi^0$ 
($\gamma^*$ is a virtual photon), is typically considered as the main 
correction to IA, necessary, in particular, to improve the description of 
the SF $A(Q^2)$ \cite{Ar75}. However the relative role of MEC is strongly model 
dependent \cite{Bu92} as the coupling constants for meson-NN-vertexes 
are not well known and arbitrary form factors are often added \cite{Ad64,VO95}.

It should be pointed out that the $\gamma^*\pi\rho$ vertex is of 
magnetic nature and its contribution to $A(Q^2)$ has to be of the same order of 
magnitude as the relativistic corrections.

 The forthcoming data about 
\gen, planned at JLab  up to $Q^2$=2 (GeV/c)$^2$, \cite{Madey}
will be crucial in this respect. The large sensitivity of the 
deuteron structure to the nucleon form factors 
shows the necessity to reconsider the role of meson exchange currents
in the deuteron physics at large momentum transfer.

\subsection{The nuclear structure}

The measurement of polarization transfer for inelastic $\vec p$ and $\vec d$ 
scattering by nuclei allows the study of the nuclear response and to disentangle 
the spin 
and isospin components. In particular it has been shown that he spin-flip 
probability is a very good signature of the presence of $\Delta S=1$ in the 
continuum, as well as for discrete states.
A systematic work has been carried on at LAMPF and at SATURNE, on several 
nuclei, from $^{12}C$ to $^{208}Pb$. The nuclear response has been measured in 
different channels and compared to RPA calculations, in order to learn about 
correlations and collectivity in nuclei \cite{Ba97}.

In order to go further, one should complete measurements at 
very small angles (with a zero degree facility) and with high statistics, and  
make a 
multipole analysis in the continuum.

\subsection{The baryon resonances excitation in the inclusive \lowercase{$\vec 
d,p$} scattering}

An application of this method based on the measurement of the spin-flip 
probability, is the 
study of nucleon resonances, in the reaction $\vec d+p\to \vec d +X$. The 
selectivity of reactions such as $p(d,d')X$
or $p(\alpha,\alpha ')X$ \cite{Mo92} to the isoscalar part of the
$N^*$-electroexcitation makes these processes complementary to
electron-nucleon inelastic scattering, for the study of the $N^*$-structure. 
Particular attention has been devoted to the Roper resonance \cite{Ro64}
In case of polarized deuteron beam, it has been shown \cite{Re96} that
the $\omega$-exchange model gives a natural and simple description of 
the polarization phenomena for $\vec d+p\rightarrow d+X$. The main ingredients 
of such model are the  existing information about
the deuteron electromagnetic form factors \cite{Ch88} and the ratio $r$ of the 
longitudinal and transversal isoscalar cross sections for the excitation of the 
$N^*$-resonances  \cite{Bi99}. 

The $\omega$-meson is preferred, among the isoscalar 
mesons as $\sigma$ or $\eta$, for several reasons. The $\omega NN-$ coupling is 
large; the $\omega$-meson, being a spin 1 particle,  can induce strong 
polarization 
effects and an energy-independent 
cross section. When it is considered as an $isoscalar~photon$, then the cross 
sections and the 
polarization observables can be calculated from the known electromagnetic 
properties of the deuteron and $N^*$, through 
the vector dominance model.

The tensor analyzing power in  $d+p\rightarrow d+X$, $T_{20}$, can be written 
in terms of the deuteron electromagnetic form factors as:
\begin{equation}
T_{20}=-\sqrt{2}\frac{V_1^2+(2V_0V_2+V_2^2)r(t)}
{4V_1^2+(3V_0^2+V_2^2+2V_0V_2)
r(t)},\label{li1}
\end{equation}
where  $V_0(t)$, $V_1(t)$ and $V_2(t)$ are linear combinations of the standard 
electric, $G_c$, magnetic $G_m$ and quadrupole $G_q$ deuteron 
form factors.
The ratio $r$ characterizes the relative role of longitudinal and transversal 
isoscalar 
excitations in the transition $\omega+N\rightarrow X$.

From Eq. (5) one can see that all information about the $\omega N N^*$-vertex 
is 
contained in the function $r$ only.
A zero value of $r$ results in a $t-$ and $w$-independent value for 
$T_{20}$, namely $T_{20}=-1/\sqrt{8}$, for any value of the deuteron 
electromagnetic form factors. The ratio $r$ is calculated using the 
collective string model in \cite{Bi99},
assuming $SU_{sf}(6)$ symmetry, including the contributions of the following 
resonances: $ N_{11}(1440),~S_{11}(1535),D_{13}(1520)~\mbox{ and} 
~S_{11}(1650)$,
which are overlapping in this energy region.

In Fig. 5 we report the theoretical predictions for $T_{20}$, in framework of 
$\omega$-model, together with the existing 
experimental data. In such approximation $T_{20}$ is a universal function of 
$t$ 
only, without 
any dependence on the initial deuteron momentum. The experimental values of 
$T_{20}$ for 
$p(\vec d,d)X$ \cite{lns250,Az96}, for different momenta of the incident 
beam are shown as 
open symbols. These data show a scaling as a function of $t$, with a small 
dependence on the 
incident momentum, in the interval 3.7-9 GeV/c. On the same plot the data for 
the elastic 
scattering process $e^-+d\rightarrow e^-+d$ \cite{t20} are shown (filled 
stars).

These different data show a very similar behavior: negative values, with a 
minimum in the region $|t|\simeq  0.35~GeV^2$ and their value increase toward 
zero at larger  $|t|$. The lines are the 
result of the $\omega$-exchange model for the $d+p\rightarrow d+X$ process: for 
$r=0$ (dashed-dotted line), the calculation based on \cite{Bi99} 
for the  Roper excitation only is represented by 
the dotted line and for the 
excitation of all the four resonances by the the full line.
The deuteron electromagnetic form factors have been taken from \cite{Ch88}, a 
calculation based 
on relativistic impulse 
approximation, and they reproduce well the $T_{20}-$data for $ed$ elastic 
scattering 
\cite{t20}.
When $r\gg 0$ or if the contribution of the deuteron magnetic form factor 
$V_1(t)$ is 
neglected, then $T_{20}$ does not depend on the ratio $r$, and coincides with 
$t_{20}$ for 
the elastic $ed$-scattering (with the same approximation).

From Fig. 5 it appears that the $t-$behavior of $T_{20}$ is very sensitive to 
the value of 
$r$ especially at relatively small $r$, $r\leq 0.5$. The values of $r$, 
predicted by 
model 
\cite{Bi99}, give a very good description of the data, when taking into 
account 
the 
contribution of all four resonances. These data, in any case, exclude a 
very 
small value of 
$r,~r \ll 0.1$ as well as very large values of $r$. 
Only the Roper resonance has a nonzero isoscalar longitudinal form factor.
Without excitation of the Roper resonance,  $r=0$, (in the considered 
kinematical region) and
the value for
$T_{20}$ becomes $t-$independent: $T_{20}=-1/2\sqrt{2}$, in evident
disagreement with existing data.

Due to their specific quark structure, the resonances lying in the concerned 
mass 
region, such as $S_{11}(1535)$,
$D_{13}(1520)$  and $S_{11}(1650)$, are characterized by a pure isovector 
nature 
of 
longitudinal virtual photons absorbed by the nucleons. 
The \underline{isoscalar} longitudinal amplitudes of
$S_{11}(1535)$ and $D_{13}(1520)$ electroexcitation vanish due to a specific  
spin-flavor symmetry,
while both isoscalar and isovector longitudinal couplings of
$S_{11}(1650), D_{15}(1675)$ and $D_{13}(1700)$ vanish identically. 

This behavior of the isoscalar form factors is essential for the correct
description of the existing experimental data on the $t-$dependence of
$T_{20}$ for the process $d+p\rightarrow d+X$.

One could improve  the model taking into 
account for example, other meson exchanges,
or the effects of the strong interaction in initial and final states. However 
these corrections are strongly model- and parameter- dependent and the existing
experimental data are not
good enough to constrain the additional parameters which have to be added. 

The successful description of the
polarization observable $T_{20}$ can be considered as a strong  indication that
the $\omega-$ exchange is the main mechanism for the considered process  and 
that 
the Roper resonance is excited in this process.

\section{Conclusions}
Some of the recent  results obtained in the study of the deuteron structure
have been reviewed. Despite the large precision and the new region of internal 
momentum explored, questions arised many decades ago are still actual:
\begin{itemize}
\item The relative role of different possible components in the deuteron wave 
function: isobar configurations, six-quark components, etc..;
\item The relativistic description of the deuteron structure: the number of 
independent components of the deuteron wave functions as well as  the number 
and 
the nature of their arguments;
\item The relative role of the different possible mechanisms in the simplest 
reactions with deuterons: for example the role of meson exchange current in the 
description of the electromagnetic form factors of the deuteron;
\item The kinematical region corresponding to the transition regime in the 
deuteron structure: i.e. from the 
nucleon-meson picture of the deuteron to the quark-gluon description.
\end{itemize}
These problems are far from being solved. The experimental study of 
polarization 
phenomena remains the most fruitful way in this investigation. If, in the case 
of 
elastic $ed$-scattering we are near the limits of the experimental 
possibilities, 
in reactions such as the deuteron photodisintegration, $\gamma+d\to n+p$ or 
the coherent pion photoproduction on on the deuteron $\gamma+d\to d+\pi^0$, a 
wide 
program of polarization experiments can be realized, due to the rich spin 
structure of the corresponding matrix elements. Moreover the experimental study 
of 
new reactions, such as the coherent neutral pion production on the deuteron, 
$e+d\to e+d+\pi^0$, in the near threshold region and in the $\Delta$-region 
(for 
large momentum transfer) can bring useful information.

These conclusions apply to the hadronic sector, too. The complete experiment for 
the simplest case, the $dp$ backward elastic scattering, requires further  steps 
which can 
be realized in principle in the nearest future. 
The study of the nucleon resonances through reactions induced by isoscalar 
probes 
as deuterons and $\alpha$ particles seems to be very promising, in particular 
the 
polarization observables.

\begin{figure}
\begin{center}
\mbox{\epsfxsize=8.cm\leavevmode \epsffile{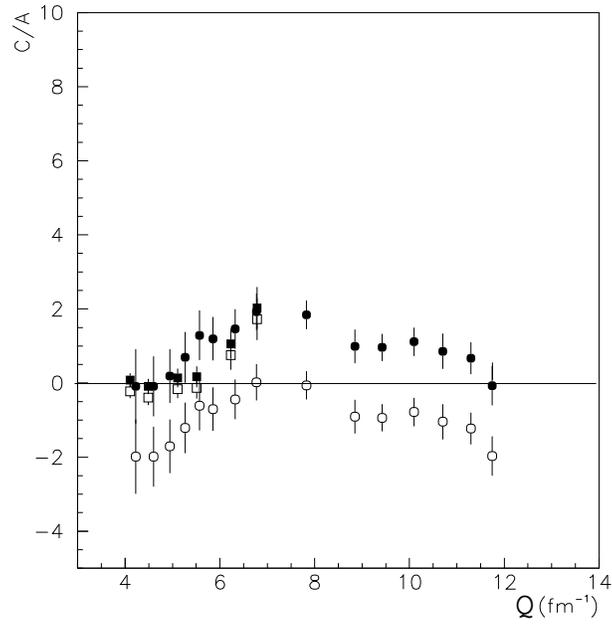}}
\end{center}

\caption{ Ratio C/A for the two sets of data, as a function of $Q$ (fm$^{-1}$):
 squares from \protect\cite{Ab99};  circles from \protect\cite{Al99}, 
corresponding solid 
symbols after renormalization (see text) }
\label{fig1}
\end{figure}
\begin{figure}
\begin{center}
\mbox{\epsfxsize=8.cm\leavevmode \epsffile{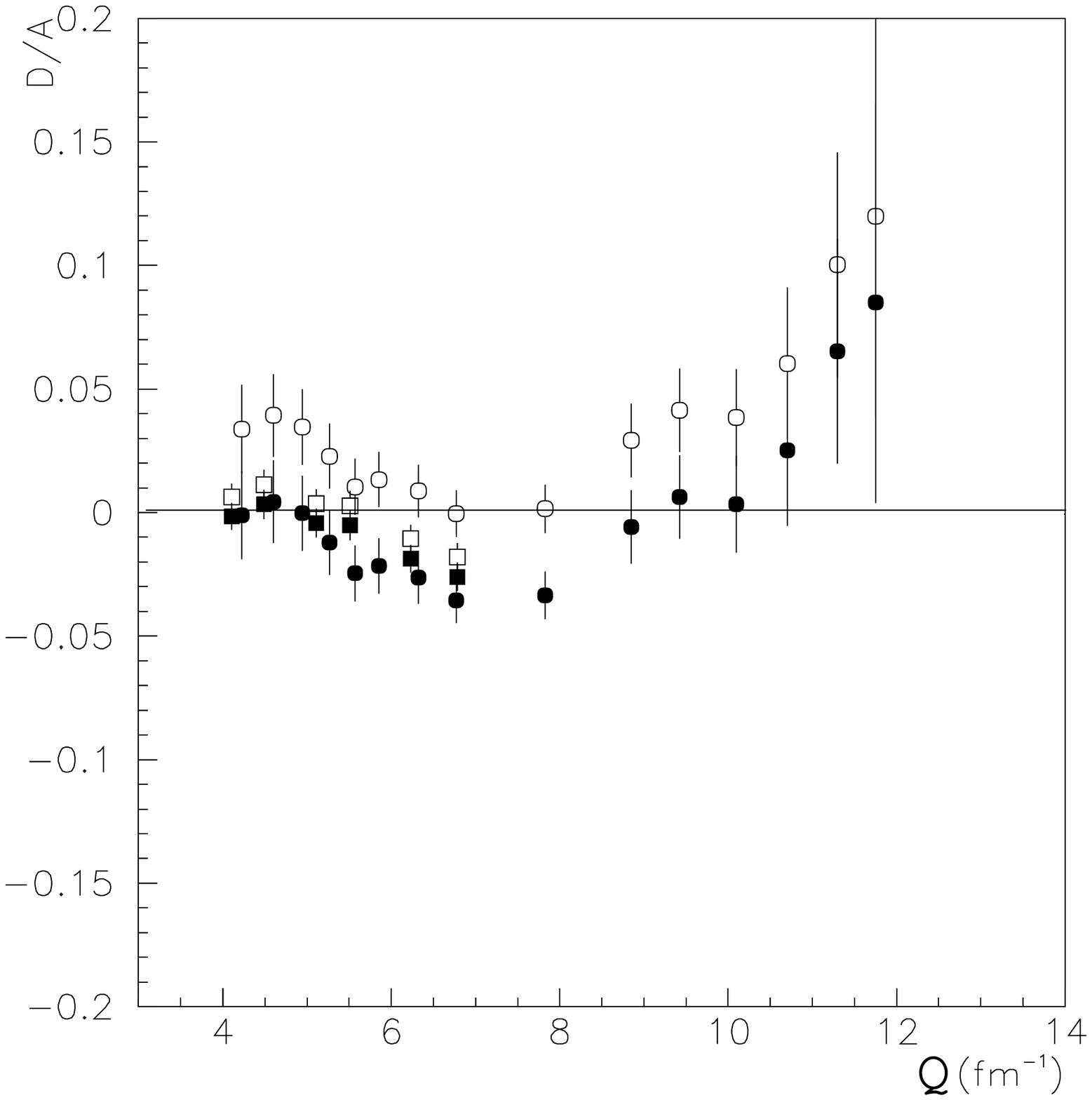}}
\end{center}
\caption{ Ratio D/A for the two sets of data, as a function of $Q$ (fm$^{-1}$): 
squares 
from \protect\cite{Ab99};  circles from \protect\cite{Al99}, corresponding 
solid 
symbols after 
renormalization (see text) }
\label{fig2}
\end{figure}

\begin{figure}
\begin{center}
\mbox{\epsfxsize=14.cm\leavevmode \epsffile{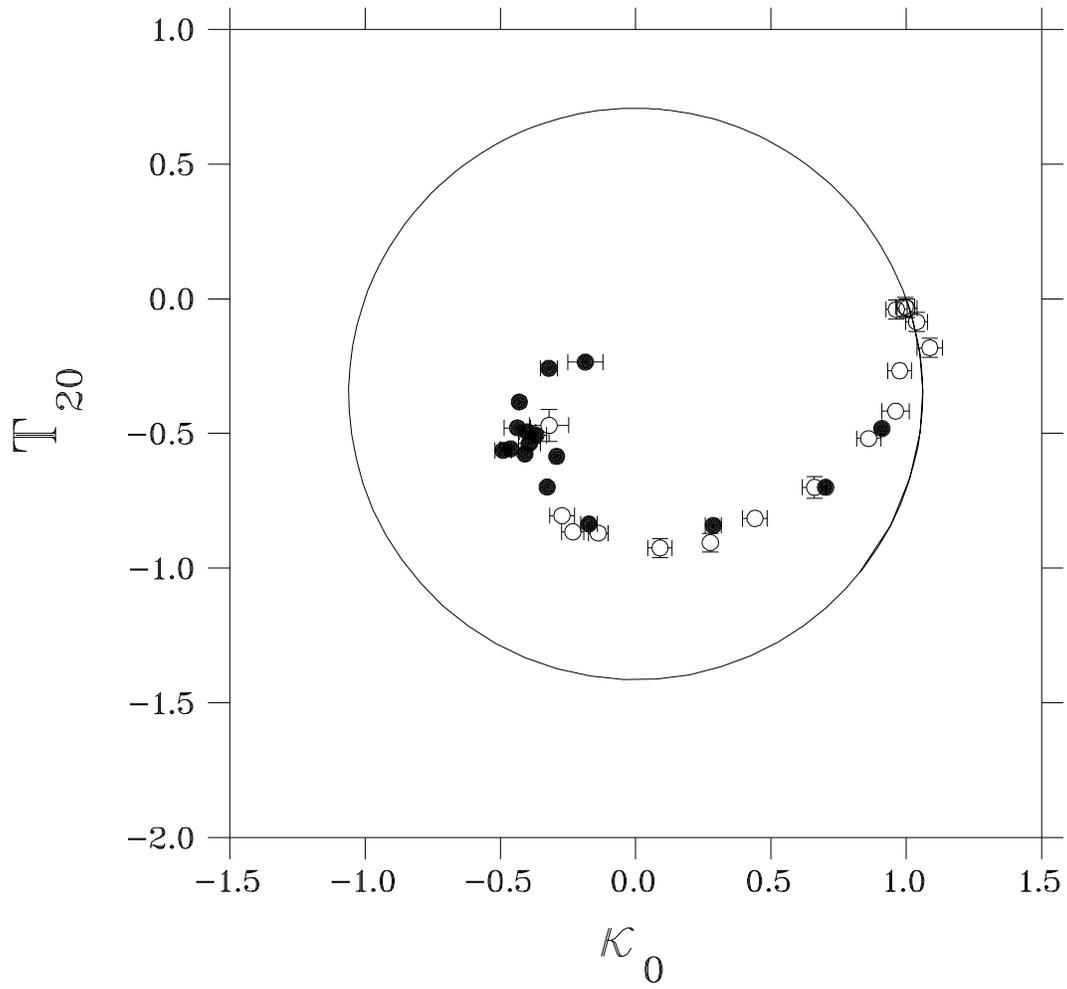}}
\end{center}
\caption{ Tensor analyzing power versus polarization transfer coefficient for 
$dp$ backward elastic (filled circles) and for inclusive break-up (open 
circles). The solid curve is the IA prediction. }
\label{fig3}
\end{figure}

\begin{figure}
\begin{center}
\mbox{\epsfxsize=10.cm\leavevmode \epsffile{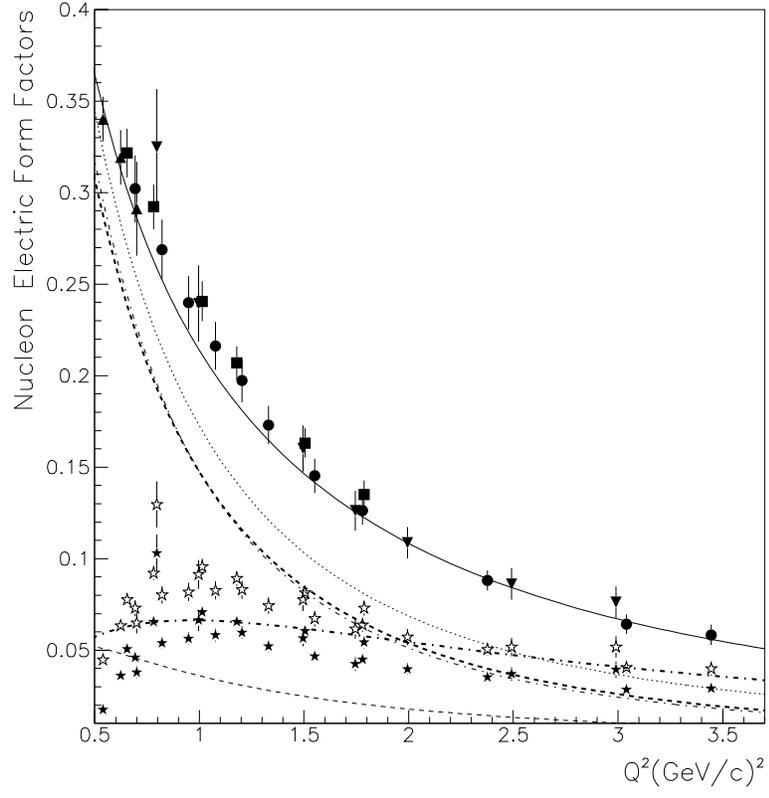}}
\end{center}
\caption{Nucleon  electric form factors as functions of 
the momentum transfer $Q^2$. in the framework of IA with Paris potential.
The isoscalar electric  form factor is derived from  the deuteron elastic 
scattering data: \protect\cite{Pl90} (solid triangles), \protect\cite{Al99} 
(solid circles), \protect\cite{Ab99} (solid squares), and \protect\cite{Ar75} 
(solid reversed triangles). 
The  electric neutron form 
factor is shown as solid stars when calculated from the dipole
representation of \gep\  (dotted line) and open stars when 
Eq. (5) is taken for \gep\  (thin dashed-dotted line).  The parametrization  
\protect\cite{G-K} is shown for \ges (solid line), for \gen\  (thick 
dashed-dotted 
line) and for \gep\  (thick dashed line). The thin dashed line is the 
parametrization \protect\cite{Galster} for \gen.}
 \label{fig4}
\end{figure}

\begin{figure}
\begin{center}
\mbox{\epsfxsize=8.cm\leavevmode \epsffile{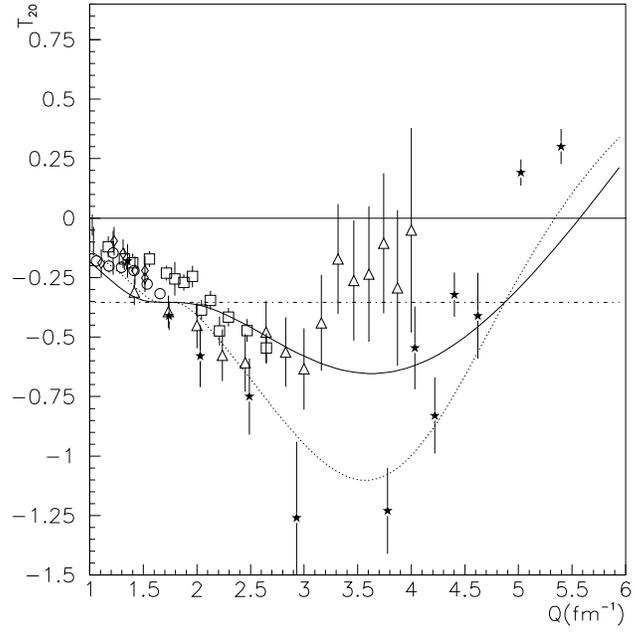}}
\end{center}
\caption{Experimental data for $T_{20}$ for
$e^-+d\rightarrow e^-+d$ elastic scattering (filled stars) \protect\cite{t20}
 and
$ d+p\rightarrow d +X$ at incident momenta of 3.75 GeV/c
 (open diamonds) \protect\cite{lns250}, 5.5 GeV/c (open circles),
4.5 GeV/c  (open squares),  9 GeV/c
 (open triangles) {\protect\cite{Az96}}.
Prediction of the $\omega-$exchange model for  $r=0$
(dashed-dotted line).
Calculations are shown for the case when only the Roper resonance is considered
(dotted line) and for the case when all the four
resonances  (6) are considered (solid line).}
\end{figure}

\end{document}